\documentclass{appolb}
\usepackage{epsfig}

\begin{document}
\title{Nucleon-deuteron capture with chiral potentials
}
\author{R.~Skibi\'nski$^a$, J.~Golak$^a$, H.~Wita{\l}a$^a$, 
W.~Gl\"ockle$^b$, A.~Nogga$^c$, E.~Epelbaum$^{c,d}$
\address{$^a$M. Smoluchowski Institute of Physics, Jagiellonian University,
                    PL-30059 Krak\'ow, Poland,\\
$^b$Institut f\"ur Theoretische Physik II, Ruhr-Universit\"at Bochum,
D-44780 Bochum, Germany\\
$^c$Institut f\"ur Kernphysik, Forschungszentrum J\"ulich,
D-52425 J\"ulich, Germany\\
$^d$
Helmholtz-Institut f\"ur Strahlen- und Kernphysik (Theorie), Universit\"at
Bonn, Nu{\ss}allee 14-16, D-53115 Bonn, Germany.}
}
\maketitle
\begin{abstract}
Present day chiral nucleon-nucleon potentials up to next-to-next-to-next-to 
leading order and three nucleon forces at next-to-next-to leading order
are used to analyze nucleon-deuteron radiative capture at
deuteron laboratory energies below E$_d\approx$ 100 MeV.
The differential cross section and the deuteron analyzing powers 
$A_y(d)$ and $A_{yy}$ are
presented and compared to data.
The theoretical predictions are obtained in the momentum-space Faddeev 
approach using the nuclear electromagnetic current operator with  
exchange currents introduced 
via the Siegert theorem.
The chiral forces provide the same quality of data description as 
a combination of the two-nucleon  AV18 and the three-nucleon 
Urbana IX interactions. However, the
different parametrizations of the chiral potentials lead to broad bands 
of predictions. 
\end{abstract}
  
\section{Introduction}

Various modern (semi)phenomenological nucleon-nucleon (NN) potentials
~\cite{stokes_nij,wiringa_AV18,machleidt_CDBonn}
adjusted to the nucleon-nucleon data allow for a
relatively good description of few-nucleon systems.
Moreover, the existing models of a three nucleon force (3NF) 
when combined with those potentials led generally to an 
 improvement in the description of three and four 
nucleon bound states~\cite{carlson,pieper,nogga00,nogga02} and reactions in 
the 3N-continuum~\cite{3nrep,3ncel,3ncbr}.
Despite indisputable successes such forces lack 
reliable theoretical justification  in the underlying theory of
strong interactions - the quantum chromodynamics (QCD). 
Recently, a new family of nuclear potentials has 
been derived~\cite{epelbaum_review,bedaque02,entem,epelbaum2005} which has 
a direct connection to QCD. 
Incorporating the spontaneously and explicitly broken chiral symmetry,
which plays a fundamental role in QCD, it is possible to construct 
nuclear forces in the framework of chiral perturbation theory. 
This approach has been originally proposed by Weinberg~\cite{weinberg} and
further developed by Ord\'o\~nez et al.~\cite{ordonez}, Kaiser et al.~\cite{kaiser},
Entem et al.~\cite{entem},  
and by Epelbaum et al.~\cite{epelbaum2005, epelbaum2004}.
(See also recent review articles~\cite{beane, bedaque02, epelbaum_review} and references
therein). Up to now, NN and 3N forces have been worked out and applied in
few-nucleon systems up to next-to-next-to-next-to-leading order (N$^3$LO)
and next-to-next-to-leading order (N$^2$LO), respectively. The leading
4N-forces have also been worked out recently~\cite{epelbaum06}.
The chiral forces have already been successfully used to describe
the elastic nucleon-deuteron scattering~\cite{epelbaum2002} and
 the deuteron breakup processes~\cite{epelbaum2002, kistryn}.
We remind the reader
that applications of chiral nuclear forces are limited to the low-energy
region.

Chiral effective field theory has also been used to study processes with external pions
and/or photons, see~\cite{beane03, beane05, krebs, lensky} for some recent applications.
While the neutron-proton radiative capture has already been considered in this framework~\cite{park2000}, 
no such analysis is available for electromagnetic processes with three nucleons.
In this study we would like, at least partially, to  fill this gap 
 by presenting for the first time the chiral predictions 
for the $n+d \rightarrow \gamma + ^3H$ 
and $p+d \rightarrow \gamma + ^3He$ processes.

We use the chiral NN potentials~\cite{epelbaum2005} at next-to leading order (NLO), next-to-next-to 
leading order (N$^2$LO) 
 and  at next-to-next-to-next-to leading order (N$^3$LO) 
 of chiral expansion.
In the N$^2$LO where the 3NF contributes for the first time, 
we also include such 3NF's~\cite{epelbaum2002}  consistent with the 
two-body interaction at this order.
A complete study of electromagnetic processes in the
chiral effective field theory would require the usage of the nuclear
electromagnetic current operator derived consistently with the applied
nuclear forces. Such a current operator is, however, not yet available,
see Ref.~\cite{park96} for some early work along these lines. For that reason,
we decided to adopt in the present study our standard approach~\cite{golak}, in
which the electromagnetic current operator is taken in the form of
a one-body current supplemented with some many-body contributions
introduced via the Siegert theorem.

In Section 2 we shortly describe our theoretical formalism. 
Results and a comparison to data are presented in Section 3. We summarize in 
Section 4.

\section{Theoretical formalism}
The present study is based on the formalism 
described in detail in~\cite{golak,skib2b,raport2005}.
The calculations are done for the two-body 
photodisintegration of the 3N bound state and the nuclear 
matrix element for the radiative $Nd$-capture 
$N_{\mu}^{rad}$ is found by applying the 
time reversal to the photodisintegration amplitude $N_{\mu}^{Nd}$.
The latter one is a matrix element of the electromagnetic current 
operator $j_{\mu}$ 
between the initial 3N bound state $\mid \Psi_b \rangle$ 
and the final scattering state  
$\mid \Psi_{Nd}^{(-)} \rangle$ 
\begin{equation}
N^{Nd}_{\mu} \equiv \langle \Psi_{Nd}^{(-)} \mid j_{\mu} \mid \Psi_b \rangle \;.
\end{equation}
In the Faddeev scheme $N_{\mu}^{Nd}$ can be written in the form~\cite{skib2b}
\begin{equation}
N^{Nd}_{\mu} = \langle \psi_1 \mid (1+P) j_{\mu} \mid \Psi_b \rangle\;,
\end{equation}
where $\mid \psi_1 \langle$
is a Faddeev component of the state $\mid \Psi_{Nd}^{(-)} \rangle$ and 
$P$ is a permutation operator defined as a sum of cyclical and anti-cyclical
permutations of three particles
\begin{equation}
P \equiv P_{12}P_{23} + P_{13}P_{23},
\end{equation}
where $P_{ij}$ interchanges
the i-th and j-th nucleons.

Starting from the Schr\"odinger equation for the scattering state  
$\mid \Psi_{Nd}^{(-)} \rangle$ the transition amplitude can be written as
\begin{eqnarray}
N_\mu^{Nd} =
\langle \phi_{1} \mid  ( 1 + P ) \mid
j_\mu \mid \Psi_b \rangle +
\langle \phi_{1} \mid  P \mid \tilde U \rangle ,
\label{eq:Nnew}
\end{eqnarray}
where $\mid \phi_1 \rangle$ is a product of the internal deuteron state
and a momentum eigenstate of the free nucleon-deuteron motion.
The auxiliary state $\mid \tilde U \rangle$ fulfills
the Faddeev-like equation
\begin{eqnarray}
\mid \tilde U \rangle  &=&
\left( t G_0 + \frac12 ( 1 + P ) V_4^{(1)} G_0 ( t G_0 +1) \right) ( 1 + P ) j_\mu \mid \Psi_b \rangle \nonumber \\
&+& \left( t G_0 P + \frac12 ( 1 + P ) V_4^{(1)} G_0 ( t G_0 +1) P \right)
\mid \tilde U \rangle ,
\label{eq:Utilde}
\end{eqnarray}
where $V_4^{(1)}$ is a part of the 3NF which is 
symmetrical under the exchange of nucleons 2 and 3, $G_0$ is 
the free three-nucleon propagator,  and $t_1$ is the
two-body t-operator acting in the 2-3 subspace.

Equation (\ref{eq:Utilde}) is solved in the mo\-men\-tum-spa\-ce basis
\begin{equation}
\label{dec1}
\mid p,q,\alpha_{J} \rangle = \mid p,q, (ls)j,(\lambda,\frac{1}{2})I (jI)J;
(t\frac{1}{2})T \rangle,
\end{equation}
where $p$ and  $q$ are the magnitudes of the two Jacobi momenta.
The quantum numbers $l$, $s$, and $j$ are the two-body subsystem 
orbital angular
momentum, spin and the total angular momentum,  respectively.
The orbital angular momentum $\lambda$ of the spectator nucleon together 
with its
spin $\frac{1}{2}$ couples to the total spectator angular momentum $I$.
The angular momenta $j$ and $I$ couple finally to the
total angular momentum of the 3N system $J$. Similar 
coupling $(t\frac{1}{2})T$ occurs 
for the isospin quantum numbers.

Starting from the transition matrix element $N_\mu^{Nd}$ the observables 
follow in a standard way~\cite{raport2005}.
 
\section{Results}

The chiral potentials depend on 
low energy constants (LEC) and the cut-off parameters $\tilde{\Lambda}$ 
and $\Lambda$. 
The LEC's are determined for given $\tilde{\Lambda}$ 
and $\Lambda$ from a fit to NN and 3N data. 
The $\tilde{\Lambda}$ cut-off 
appears  in the spectral 
function regularization~\cite{epelbaum2004} and 
ensures that the high-momentum components of the two-pion exchange are
explicitly excluded from the potential.
The cut-off $\Lambda$ comes from a further regularization of the 
potential as used in the Lippmann-Schwinger equation, 
which cuts off the (meaningless) contributions of 
high-momentum states in the dynamical equation.
We performed numerical calculations for a few combinations of 
$(\Lambda, \tilde{\Lambda})$ values which covers the possible ranges of the 
regularization parameters:
(450(400 at NLO), 500), (600(550 at NLO),500(600 at N$^3$LO)), (550,600), 
(450(400 at NLO),700), 
(600(550 at NLO),700) (in units of $[{\rm MeV/c}]$).

The theoretical predictions at a given chiral order provide a band 
comprising all results. The  bands presented in the following correspond 
to the NN potential at NLO, N$^2$LO, NN+3NF at N$^2$LO, and finally to the 
NN potential at N$^3$LO.

\begin{figure}[htb]
\begin{center}
\psfig{file=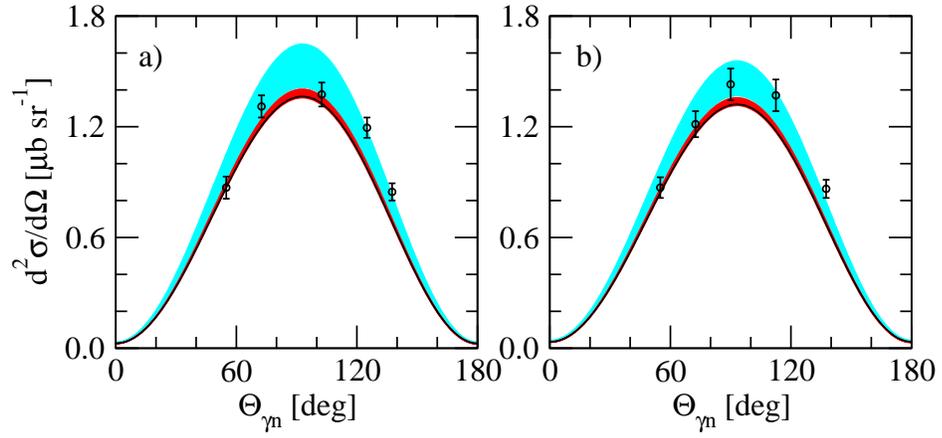,scale=0.7}
\end{center}
\caption{
The c.m. neutron-deuteron capture  cross section at:  9.0 MeV (a) and 10.8 MeV (b)
 neutron lab energies.
The light (cyan) and dark (red) shaded bands are NN NLO and NN+3NF N$^2$LO predictions, respectively.
The solid line represents prediction
obtained with the AV18 and Urbana IX forces. Data are from
\cite{mitev}.}
\label{fig1a}
\end{figure}

\begin{figure}[htb]
\begin{center}
\psfig{file=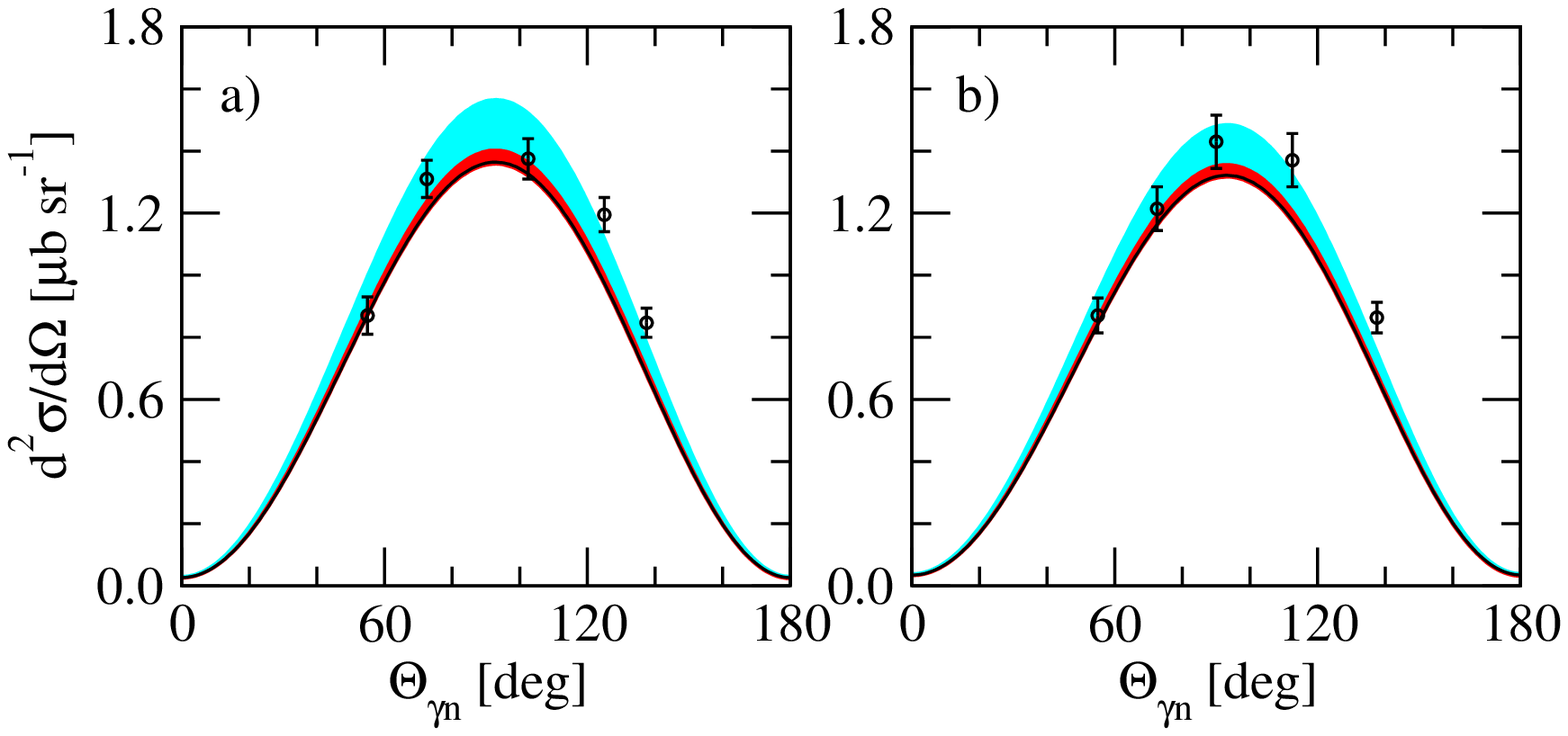,scale=0.7}
\end{center}
\caption{The c.m. neutron-deuteron capture  cross section at:  9.0 MeV (a) and 10.8 MeV (b)
 neutron lab energies.
The light (cyan) and dark (red) shaded bands are NN and NN+3NF force predictions at 
N$^2$LO. 
The line and data as in Fig.~\ref{fig1a}.}
\label{fig1}
\end{figure}

\begin{figure}[htb]
\begin{center}
\psfig{file=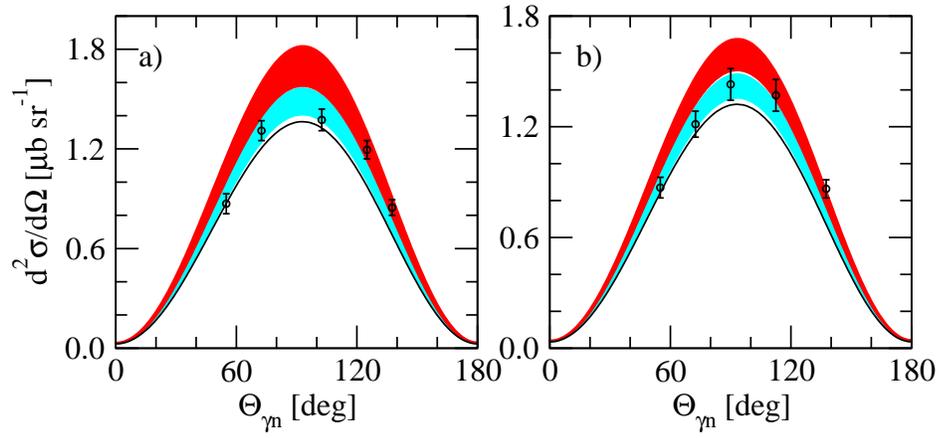,scale=0.7}
\end{center}
\caption{The c.m. neutron-deuteron capture  cross section at:  9.0  MeV
 (a) and 10.8  MeV
 (b) 
neutron lab energies.
The light (cyan) and dark (red) shaded bands are NN force predictions at N$^2$LO and N$^3$LO, respectively
The solid line and data as in Fig.~\ref{fig1a}.}
\label{fig2}
\end{figure}

\begin{figure}[htb]
\begin{center}
\psfig{file=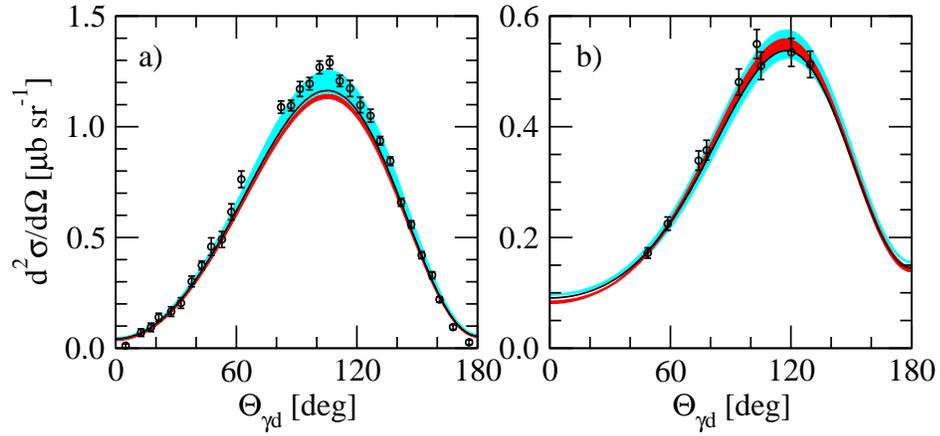,scale=0.7}
\end{center}
\caption{
The c.m. proton-deuteron capture  cross section at:  29 MeV (a) and 95.0 MeV (b)
 deuteron lab energies.
The bands and the line as in Fig.~\ref{fig1a}. 
Data at 29 MeV are from \cite{Belt} and
at 95 MeV from \cite{Pitts}.}
\label{fig3a}
\end{figure}

\begin{figure}[htb]
\begin{center}
\psfig{file=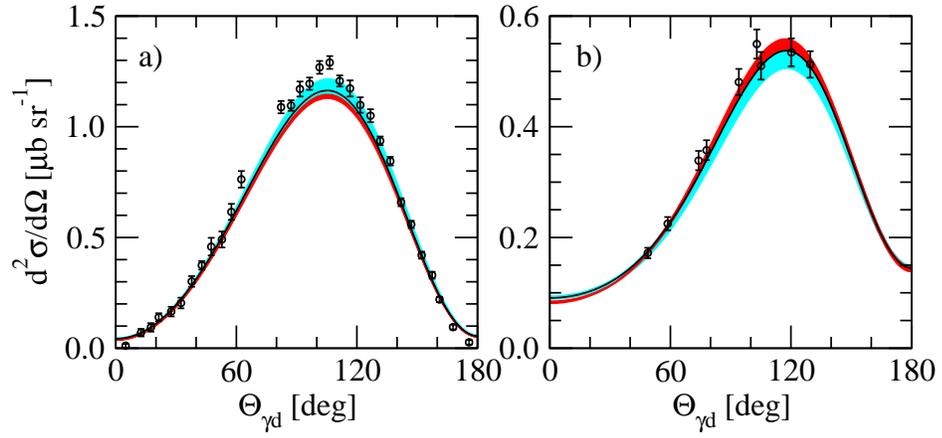,scale=0.7}
\end{center}
\caption{The c.m. proton-deuteron capture  cross sections at:  29  MeV
(a) and 95  MeV (b) 
deuteron lab. energies.
The bands and the line as in Fig.~\ref{fig1}. 
Data as in Fig.~\ref{fig3a}.}
\label{fig3}
\end{figure}

\begin{figure}[htb]
\begin{center}
\psfig{file=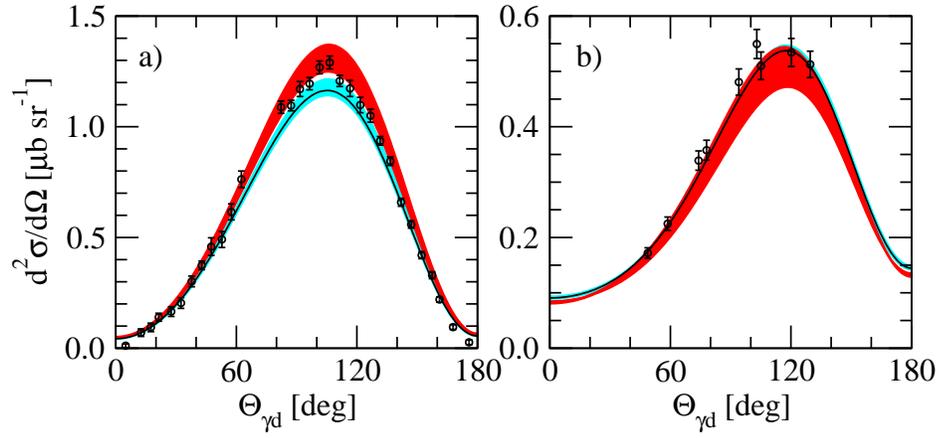,scale=0.7}
\end{center}
\caption{The c.m. proton-deuteron capture  cross sections at:  29 MeV (a) and 95 MeV (b) 
deuteron lab. energies.
The bands and the line as in Fig.~\ref{fig2}. Data as in Fig.~\ref{fig3a}.
The wider N$^3$LO band covers the N$^2$LO at E$_d$=95 MeV.}
\label{fig4}
\end{figure}

Let us start the presentation of our results with the c.m. differential 
cross section for the neutron-deuteron
(Figs.~\ref{fig1a}-\ref{fig2}) and proton-deuteron 
(Figs.~\ref{fig3a}-\ref{fig4}) capture.
In Fig.~\ref{fig1a} we display predictions at incoming  
neutron lab.  energy $E_n$=9 MeV and 10.8 MeV.
The light (cyan) band corresponds to NLO predictions 
while the dark (red) one refers to the full (i.e. NN+3NF) N$^2$LO results. 
While the NLO band is relatively broad, the N$^2$LO band is much narrower.
This shows the fast convergence of the chiral expansion for this observable.
The NN+3NF N$^2$LO chiral predictions are in
agreement with the AV18+Urbana IX combination of the standard NN and 3NF models
obtained using the same  current operator.
In Fig.~\ref{fig1} we show a scale of 3NF effects at N$^2$LO by comparison
predictions which corresponds to the full Hamiltonian (dark band)
and other ones which corresponds to the interaction restricted to the NN force
only (light band). Results obtained with NN force at N$^2$LO are similar to those
at NLO.  
Due to the width of the NN band it is difficult to conclude if the inclusion of the 3NF 
improves or not the description of the data, however the 3NF effects are clear. 
There is an improvement  at smaller $\Theta_{\gamma n}$ angles but at larger
 angles the data  
favour the pure NN predictions.  
The comparison between N$^2$LO and N$^3$LO based on NN interactions only
is shown in Fig.~\ref{fig2}.
The inclusion of the additional terms appearing in N$^3$LO (without corresponding 3NFs)
shifts the predictions 
above those of N$^2$LO and brings them 
above the data. 
It would be interesting to see if the inclusion of the 3NFs in N$^3$LO will 
decrease 
the cross section bringing theory back to the data, similarly as 
 it is the case at N$^2$LO.

In Figs.~\ref{fig3a}-\ref{fig4} proton-deuteron capture cross sections are 
shown at two incoming deuteron lab. energies, E$_d$=29 and 95 MeV. 
The nice convergence of chiral expansion is seen at both energies
(see Fig.~\ref{fig3a}). In Fig.~\ref{fig3}, 
at smaller deuteron energy E$_d$=29 MeV the inclusion of the 3NF in N$^2$LO 
shifts the results in the wrong direction, deteriorating the description of 
data. Passing to (incomplete)
N$^3$LO improves the agreement with data. At the much higher deuteron 
energy E$_d$=95 MeV the 3NF  
acts in the opposite direction, undoubtedly improving the data description.
At this energy there is no difference between predictions at 
N$^2$LO and N$^3$LO based on the NN interactions only.
However, the band of predictions at N$^3$LO is much broader 
than the corresponding one at N$^2$LO.
 
\begin{figure}[htb]
\begin{center}
\psfig{file=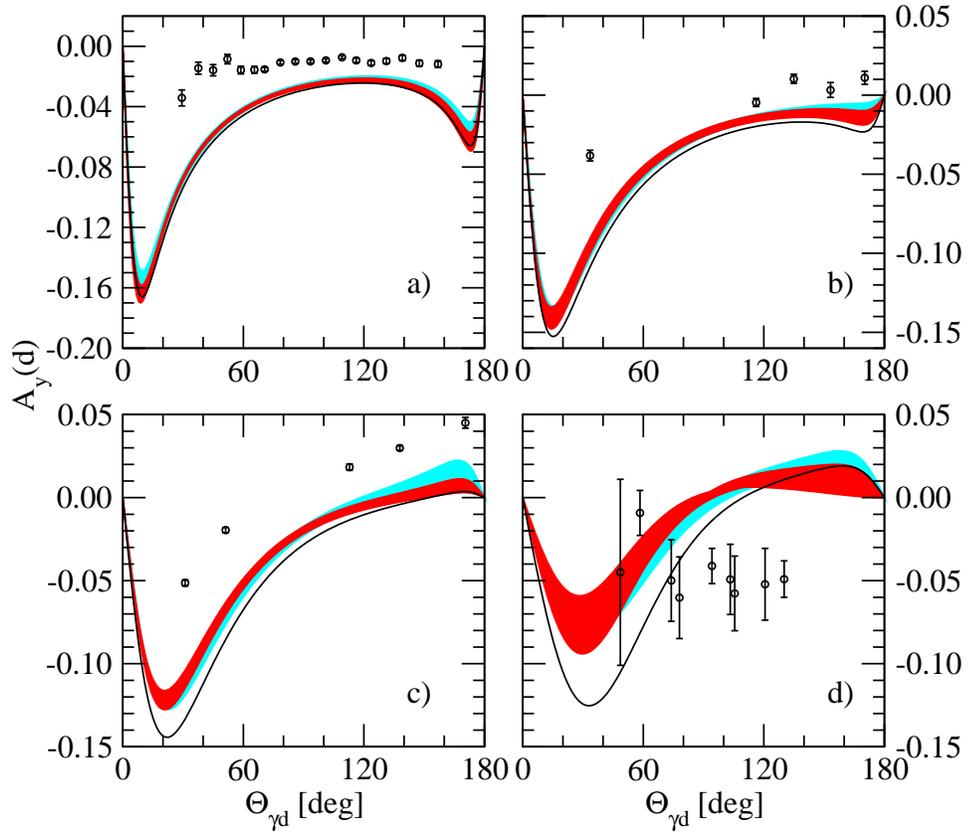,scale=0.7}
\end{center}
\caption{The deuteron vector analyzing power $A_y(d)$ at:  17.5 MeV (a),
29 MeV (b), 45 MeV (c)  and 95 MeV (d)
deuteron lab. energies.
The bands and the line as in Fig.~\ref{fig1a}. 
Data at 17.5 MeV are from \cite{Sagara.1,Sagara.2}, at 29 and 45 MeV
from \cite{Klechneva}, at 95 MeV from \cite{Pitts}.}
\label{fig5a}
\end{figure}

\begin{figure}[htb]
\begin{center}
\psfig{file=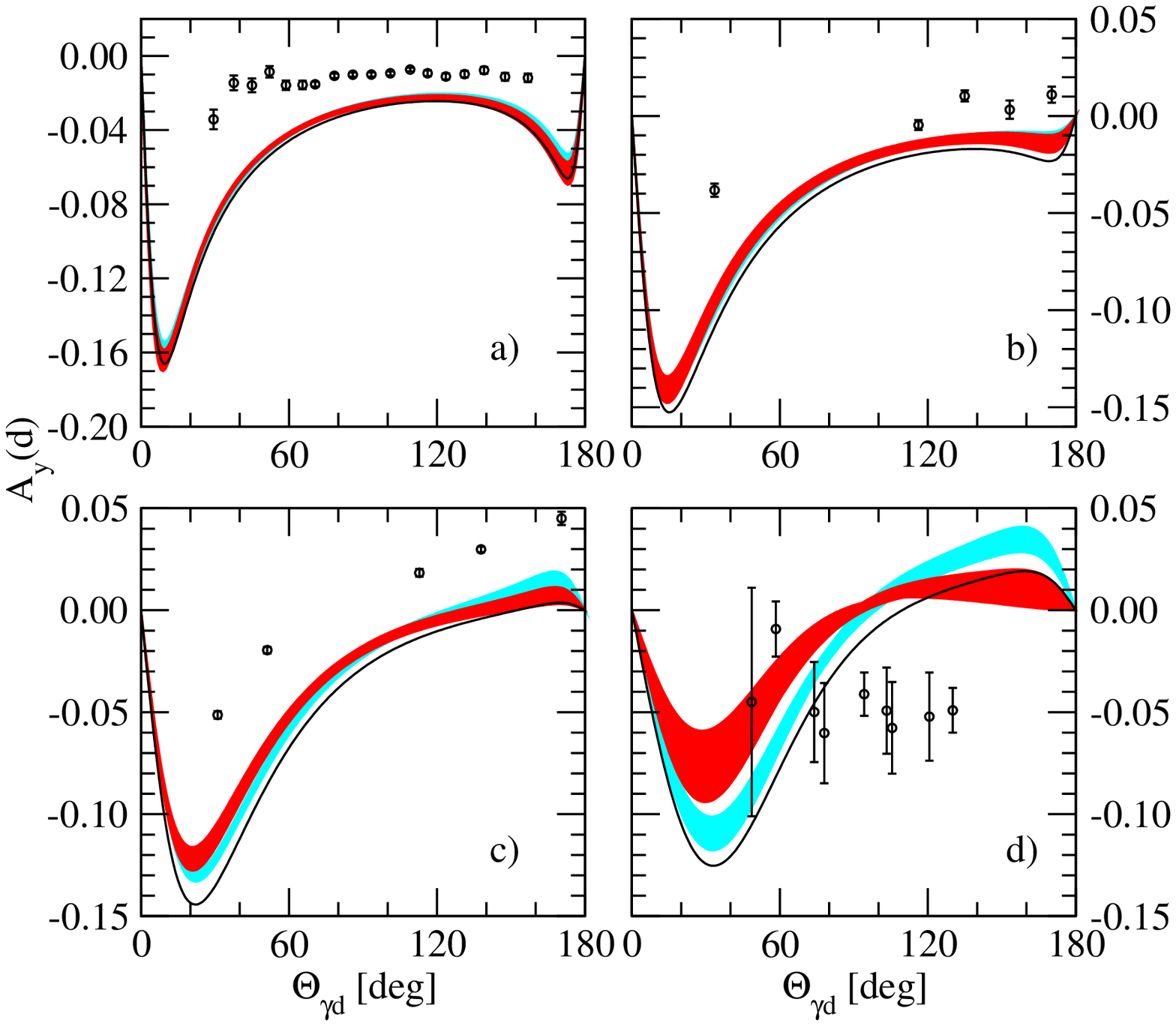,scale=0.7}
\end{center}
\caption{The deuteron vector analyzing power $A_y(d)$ at:  17.5 MeV (a),  
29 MeV (b), 45 MeV (c)  and 95 MeV (d) 
deuteron lab. energies.
The bands and the line as in Fig.~\ref{fig1}. 
Data as in Fig.~\ref{fig5a}.}
\label{fig5}
\end{figure}

\begin{figure}[htb]
\begin{center}
\psfig{file=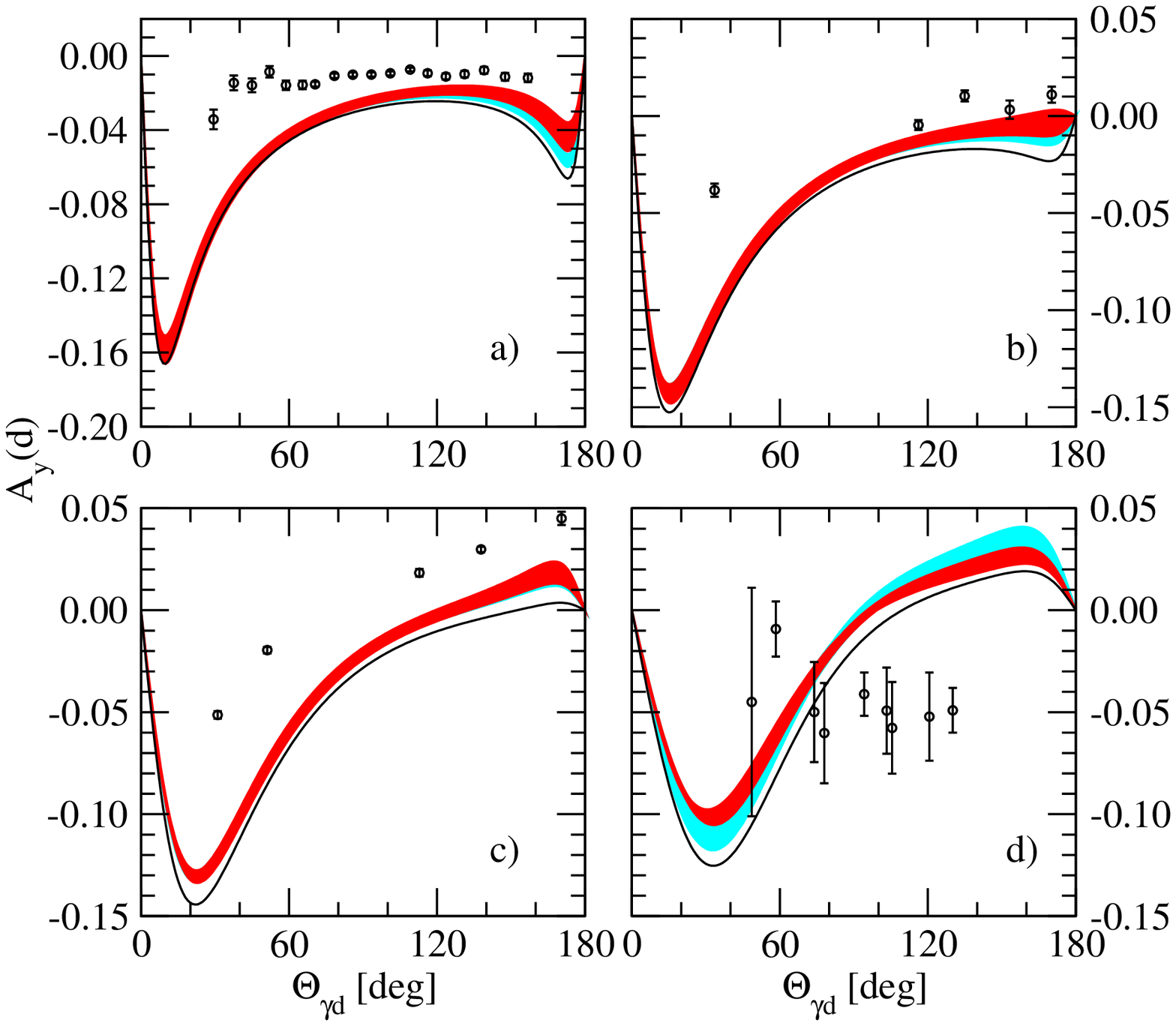,scale=0.7}
\end{center}
\caption{The deuteron vector analyzing power $A_y(d)$ at:  17.5 MeV (a), 
 29 MeV (b), 45 MeV (c)  and 95 MeV (d) 
deuteron lab. energies.
The bands and the line as in Fig.~\ref{fig2}. Data as in Fig.~\ref{fig5a}.}
\label{fig6}
\end{figure}

\begin{figure}[htb]
\begin{center}
\psfig{file=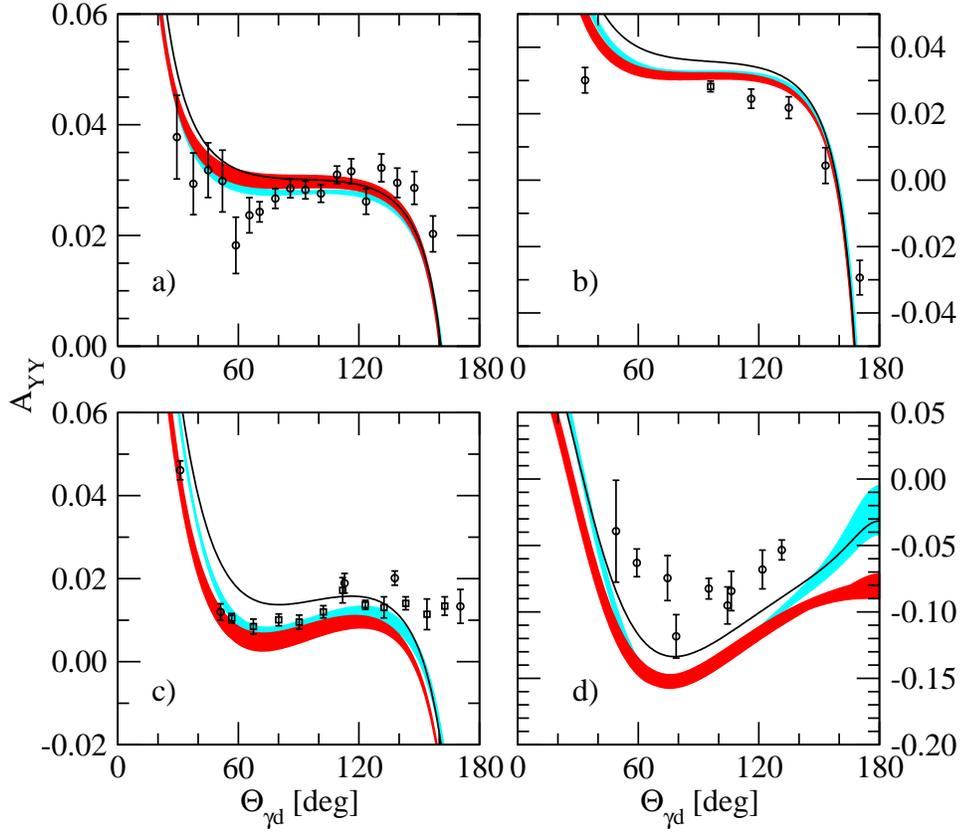,scale=0.7}
\end{center}
\caption{The deuteron tensor analyzing power $A_{YY}$ at:  17.5 MeV (a),
29 MeV (b), 45 MeV (c)  and 95 MeV (d)
deuteron lab. energies.
The bands and the line as in Fig.~\ref{fig1a}. 
Data at 17.5 MeV from \cite{Sagara.1,Sagara.2}, at 29 MeV
from \cite{Jourdan86} (squares) and \cite{Klechneva} (circles), at 45 MeV
from \cite{Klechneva} (circles) and \cite{Anklin} (squares) and at 95 MeV from \cite{Pitts}.
}
\label{fig7a}
\end{figure}

\begin{figure}[htb]
\begin{center}
\psfig{file=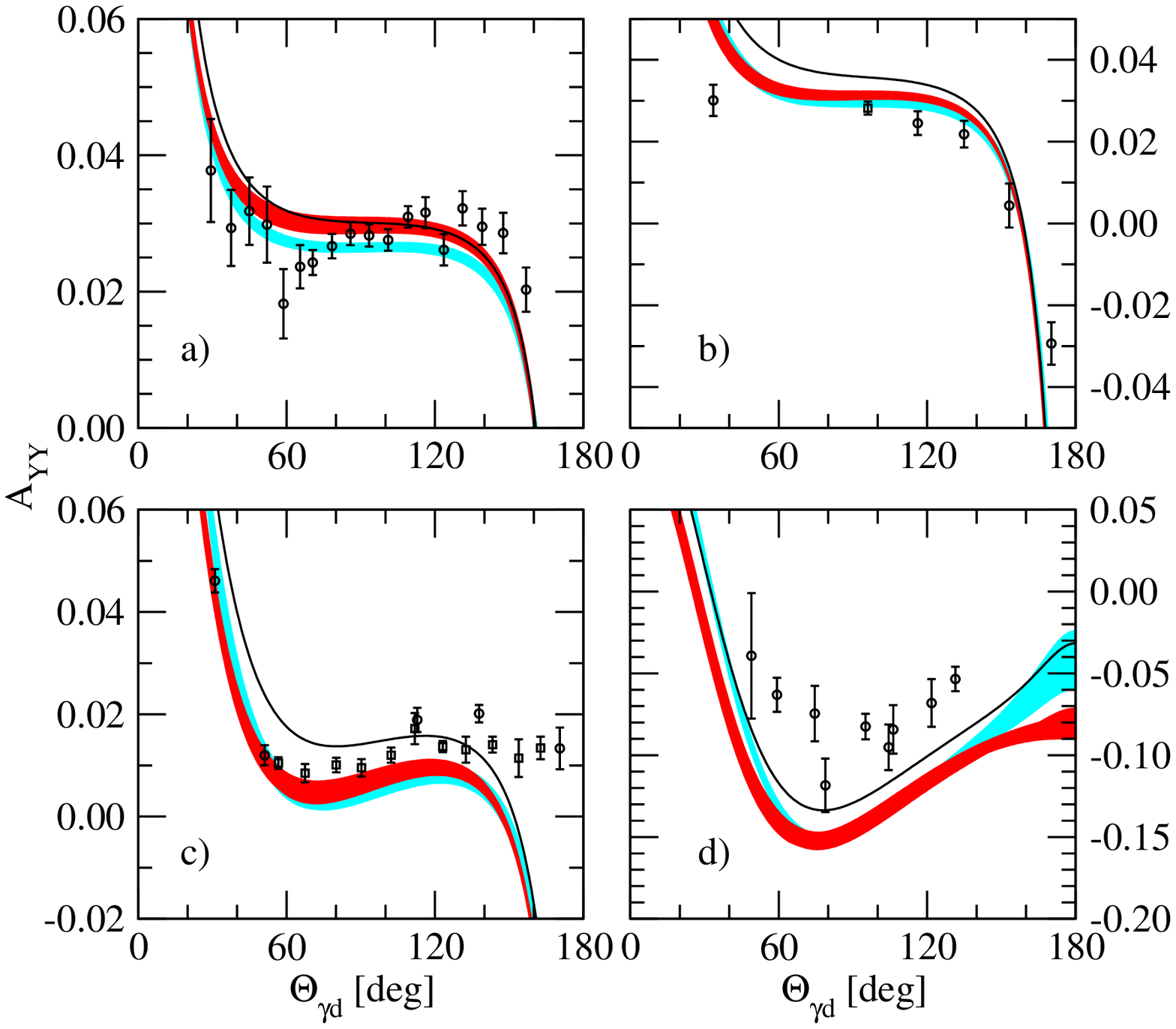,scale=0.7}
\end{center}
\caption{The deuteron tensor analyzing power $A_{YY}$ at:  17.5 MeV (a), 
29 MeV (b), 45 MeV (c)  and 95 MeV (d) 
deuteron lab. energies.
The bands and the line as in Fig.~\ref{fig1}. 
Data as in Fig.~\ref{fig7a}.}
\label{fig7}
\end{figure}

\begin{figure}[htb]
\begin{center}
\psfig{file=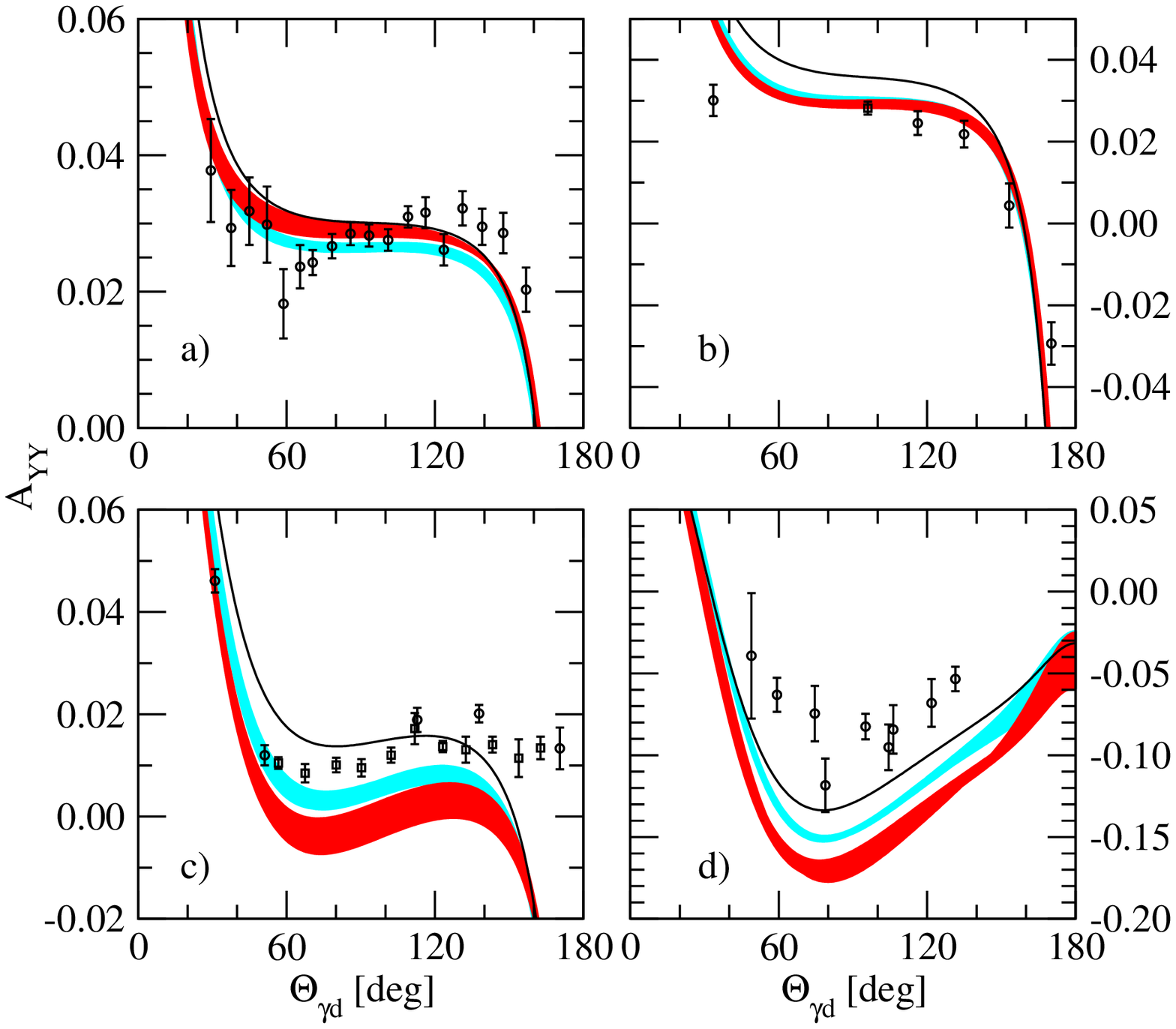,scale=0.7}
\end{center}
\caption{The deuteron tensor analyzing power $A_{YY}$ at:  17.5 MeV (a), 
29 MeV (b), 45 MeV (c)  and 95 MeV (d) 
deuteron lab. energies.
The bands and the line as in Fig.~\ref{fig2}. Data as in Fig.~\ref{fig7a}.}
\label{fig8}
\end{figure}

Now let us discuss selected spin observables. We would like to 
show results for 
the deuteron vector analyzing power $A_y(d)$ 
(Figs.~\ref{fig5a}-\ref{fig6}) and 
the deuteron tensor analyzing power $A_{yy}$ (Figs.~\ref{fig7a}-\ref{fig8})
at four deuteron energies 17.5 MeV, 29 MeV, 45 MeV and 95 MeV, 
for which comparison 
with data is possible.
Here, the convergence of the chiral expansion is much slower
(see Figs.~\ref{fig5a} and~\ref{fig7a}) and 
the bands have more or less the same width. At the highest 
deuteron energy E$_d$=95 MeV for $A_y(d)$ the N$^2$LO band is even wider 
than that one in NLO. 
The effects of the 3NF at N$^2$LO on $A_y(d)$ are not substantial 
and are seen mainly at the
highest deuteron energy. The observed discrepancy with the 
data might at least partially be
caused by the simplified model of the electromagnetic operator used 
in the present calculations.
(The results  based on the AV18+Urbana IX potential and a
more realistic model of the current
operator is shown in ~\cite{raport2005} where a better agreement with data 
was found.)
The effects of the new terms occurring in the NN potential in N$^3$LO are small,
as seen in Fig.~\ref{fig6}.
At the three lower energies they slightly shift predictions in the data direction.
At the highest energy the picture for $A_y(d)$ is similar 
when including the 3NF or the new N$^3$LO terms. 
However, in the first case the band of predictions is much broader.  

Similarly to the $A_y(d)$ case, the effects of moving from NLO to N$^2$LO, 
as well as the 3NF effects for $A_{yy}$ 
(Figs.~\ref{fig7a}-\ref{fig7}) are not significant.
The discrepancy between AV18+Urbana IX and the chiral predictions for
the higher energies of 45 MeV and 95 MeV is particularly interesting. 
At these energies the step to the next order in the chiral expansion 
seems to be necessary (see Fig.~\ref{fig8}). Moreover, 
the additional terms in the N$^3$LO NN chiral potential 
shift predictions farther away from the data. This points to the need of 
developing the corresponding 3N forces, which is under way.

\section{Summary}
The  first predictions for radiative nucleon-deuteron capture  based 
on chiral forces 
are presented. The differential cross section and the deuteron 
vector and tensor
analyzing powers have been calculated using the chiral NN potential
at NLO, N$^2$LO and N$^3$LO. Also the
results at N$^2$LO  
with the chiral  3NF of that order are shown. 
The nuclear current operator contains, besides the single nucleon 
current, some many-body contributions as introduced via the Siegert theorem.
The chiral potentials give similar predictions as
the AV18+Urbana IX combination of the standard nuclear forces. 
We found, however, a significant
theoretical uncertainty for all considered observables. The bands
of predictions are in some cases  
too broad, especially in the N$^3$LO NN and N$^2$LO NN+3NF cases, to make a 
decisive conclusion about the quality of  the Nd-capture data description. 
This broadness can be partially caused by the inconsistent current operator 
used in this study. In future studies the nuclear current
constructed in the same framework as chiral forces should be used. 
This first study indicates that the inclusion of the consistent current operator
and going to N3LO in the chiral expansion might be necessary  to obtain a
more precise description of radiative processes within the chiral effective field
theory framework.
However, we have shown that already now this framework can be
used for analyzing low energy electromagnetic processes.

\section*{Acknowledgments}
This work has been supported by the Polish Committee 
for Scientific Research under Grant No. 2P03B00825
and by the Helmholtz Association, contract number VH-NG-222.
The numerical calculations have been performed on the the IBM Regatta p690+ of the
NIC in J\"ulich, Germany.

\end{document}